\documentclass[11pt,twoside]{article}
\usepackage{cspm2015}

\aspSuppressVolSlug
\resetcounters

\def\figspath{.} 

\begin{document}

\title{The Spanish Space Weather Service SeNMEs. A case study on the Sun-Earth chain}

\author{J.~Palacios, $^1$ C.~Cid, $^1$ A.~Guerrero,$^1$ E.~Saiz,$^1$  Y.~Cerrato,$^1$  
M.~Rodr{\'{\i}}guez-Bouza,$^2$ I.~Rodr{\'{\i}}guez-Bilbao,$^2$ M.~Herraiz,$^{2,3}$ and G.~Rodr{\'{\i}}guez-Caderot$^2$}
 
\affil{$^1$ Universidad de Alcal{\'a} (UAH), Alcal{\'a} de Henares, Madrid, Spain; \email{judith.palacios@uah.es}}
\affil{$^2$ Universidad Complutense de Madrid (UCM), Madrid, Spain}
\affil{$^3$  Instituto de Geociencias, (UCM, CSIC), Madrid, Spain}

\paperauthor{J.~Palacios}{judith.palacios@uah.es}{}{Universidad de Alcal{\'a}, UAH}{Departamento de F{\'{\i}}sica y Matem{\'a}ticas}{Alcal{\'a} de Henares}{Madrid}{28871}{Spain}
\paperauthor{C.~Cid}{consuelo.cid@uah.es}{}{Universidad de Alcal{\'a}, UAH}{Departamento de F{\'{\i}}sica y Matem{\'a}ticas}{Alcal{\'a} de Henares}{Madrid}{28871}{Spain}
\paperauthor{A.~Guerrero}{antonio.guerreroo@uah.es}{}{Universidad de Alcal{\'a}, UAH}{Departamento de F{\'{\i}}sica y Matem{\'a}ticas}{Alcal{\'a} de Henares}{Madrid}{28871}{Spain}
\paperauthor{E.~Saiz}{elena.saiz@uah.es}{}{Universidad de Alcal{\'a}, UAH}{Departamento de F{\'{\i}}sica y Matem{\'a}ticas}{Alcal{\'a} de Henares}{Madrid}{28871}{Spain}
\paperauthor{Y.~Cerrato}{yolanda.cerrato@uah.es}{}{Universidad de Alcal{\'a}, UAH}{Departamento de F{\'{\i}}sica y Matem{\'a}ticas}{Alcal{\'a} de Henares}{Madrid}{28871}{Spain}
\paperauthor{M.~Rodr{\'{\i}}guez-Bouza}{martarb7187@gmail.com}{}{Universidad Complutense de Madrid, UCM}{Facultad de Ciencias F{\'{\i}}sicas}{Madrid}{Madrid}{28040}{Spain}
\paperauthor{I.~Rodr{\'{\i}}guez-Bilbao}{irbilbao@pdi.ucm.es}{}{Universidad Complutense de Madrid, UCM}{Facultad de Ciencias F{\'{\i}}sicas}{Madrid}{Madrid}{28040}{Spain}
\paperauthor{M.~Herraiz}{mherraiz@fis.ucm.es}{}{Universidad Complutense de Madrid, UCM}{Facultad de Ciencias F{\'{\i}}sicas}{Madrid}{Madrid}{28040}{Spain}
\paperauthor{G.~Rodr{\'{\i}}guez-Caderot}{grc@ucm.es}{}{Universidad Complutense de Madrid, UCM}{Instituto de Matem{\'a}tica Interdisciplinar, Facultad de Ciencias Mateml{\'a}ticas}{Madrid}{Madrid}{28040}{Spain}

\begin{abstract}

The Spanish Space Weather Service SeNMEs, \url{www.senmes.es}, is a portal created by the SRG-SW of the Universidad de Alcal{\'a}, Spain, to meet societal needs of near real-time space weather services.
This webpage-portal is divided in different sections to fulfill users needs about space weather effects: radio blackouts, solar energetic particle events, geomagnetic storms and presence of geomagnetically induced currents. 

In less than one year of activity, this service has released a daily report concerning the solar current status and interplanetary medium, informing about the chances of a solar perturbation to hit the Earth's environment. There are also two different forecasting tools for geomagnetic storms, and a daily ionospheric map. These tools allow us to nowcast a variety of solar eruptive events and forecast geomagnetic storms and their recovery, including a new local geomagnetic index, \emph{LDi{\~n}}, along with some specific new scaling.

 In this paper we also include a case study analysed by SeNMEs. Using different high resolution and cadence data from space-borne solar telescopes SDO, SOHO and GOES, along with ionospheric and geomagnetic data, we describe the Sun-Earth feature chain for the event. 
 
\end{abstract}

\section{Introduction}
The portal SeNMEs (acronym for the name `Servicio Nacional de Meteorolog{\'{\i}}a Espacial', Spanish National Space Weather Service) was created aiming at linking all the relevant features of the Sun-Earth chain, from the most active solar features, as flares, surges and filament eruptions in the chromosphere, to related structures observed in the corona, as CMEs. Following the chain, we also study events in the interplanetary medium, up to effects observed on Earth, as ionospheric disturbances and geomagnetic storms. The information offered is meant to provide a wide societal approach. Due to this, we have been awarded with the `Award on Knowledge Transfer from University to Society' of the Universidad de Alcal{\'a} in 2015.
This web service started its operations on Dec 15, 2014, having an important coverage by national and some international media.
 
 The SeNMEs team is composed by researchers from the Universidad de Alcal{\'a} (UAH) and Universidad Complutense de Madrid (UCM). 

There is not a subscription service available in this portal, although subscriptions to alerts are actually provided through the webpage \url{www.spaceweather.es}. 

\section{Web sections}
The main page (as shown in Fig.~\ref{cspm2015palacios1_fig1}) of the portal hosts a fundamental part of the service, which is the colourbar flags. The plots, colourbars and scales have been conceived to inform non-expert users of space weather conditions and their possible effects on technological systems and over population. They have been designed thinking on those users that can be affected by disturbed space weather conditions. The scales show the level of disturbance of terrestrial environment in four different situations: radio blocking \emph{(R)}, solar radiation storm \emph{(S)}, geomagnetic storm \emph{(G)} and geomagnetically induced currents, GICs \emph{(C)}. Every scale has different levels (shown in different colours). These levels are related to their intensity and the frequency of occurrence. Moreover, the appearance of a certain level in a scale is related to the intensity of the physical processes involved in the disturbances of the terrestrial environment. The \emph{S} and \emph{R} scales (i. e., radio disturbances due to flares, and solar particle flux, respectively) are coincident with those previously introduced by NOAA, but \emph{G} and \emph{C} scales have been specifically designed for Spain. They are based on the one-minute resolution Local Disturbance Index for Spain \emph{(LDi{\~n})} and its derivative, instead of $K_{p}$ index. Four graphs in the shape of a column with increasing intensity from green colour (lower, quiet time) to red (highest, most severe disturbance) show the situation for \emph{R}, \emph{S}, \emph{G} and \emph{C}. Inside every column, two shadowed rectangles correspond to the real-time space weather conditions (the right-most one in the column) and to those from the most recent past (the left-most one in the column). For \emph{R} and \emph{S} scales, real-time indicators show the maximum disturbance in the last 30 minutes and the most recent past corresponds to the last 2 hours. For \emph{G} and \emph{C} scales, these indicators correspond to the last 2 hours and 24 hours respectively. This is shown in Fig.~\ref{cspm2015palacios1_fig2}.

There is also an important section, that is the `Last report'. Every day we compile and analyse solar features that may trigger or have triggered an impulsive or eruptive event in the solar atmosphere (from the photosphere up to the corona). We also analyse the interplanetary medium to confirm or rule out observed solar features that can propagate all the way to the Earth. The final part is reporting the geomagnetic and ionospheric disturbances,  along with the possible formation of geomagnetically induced currents.

\articlefigure{\figspath/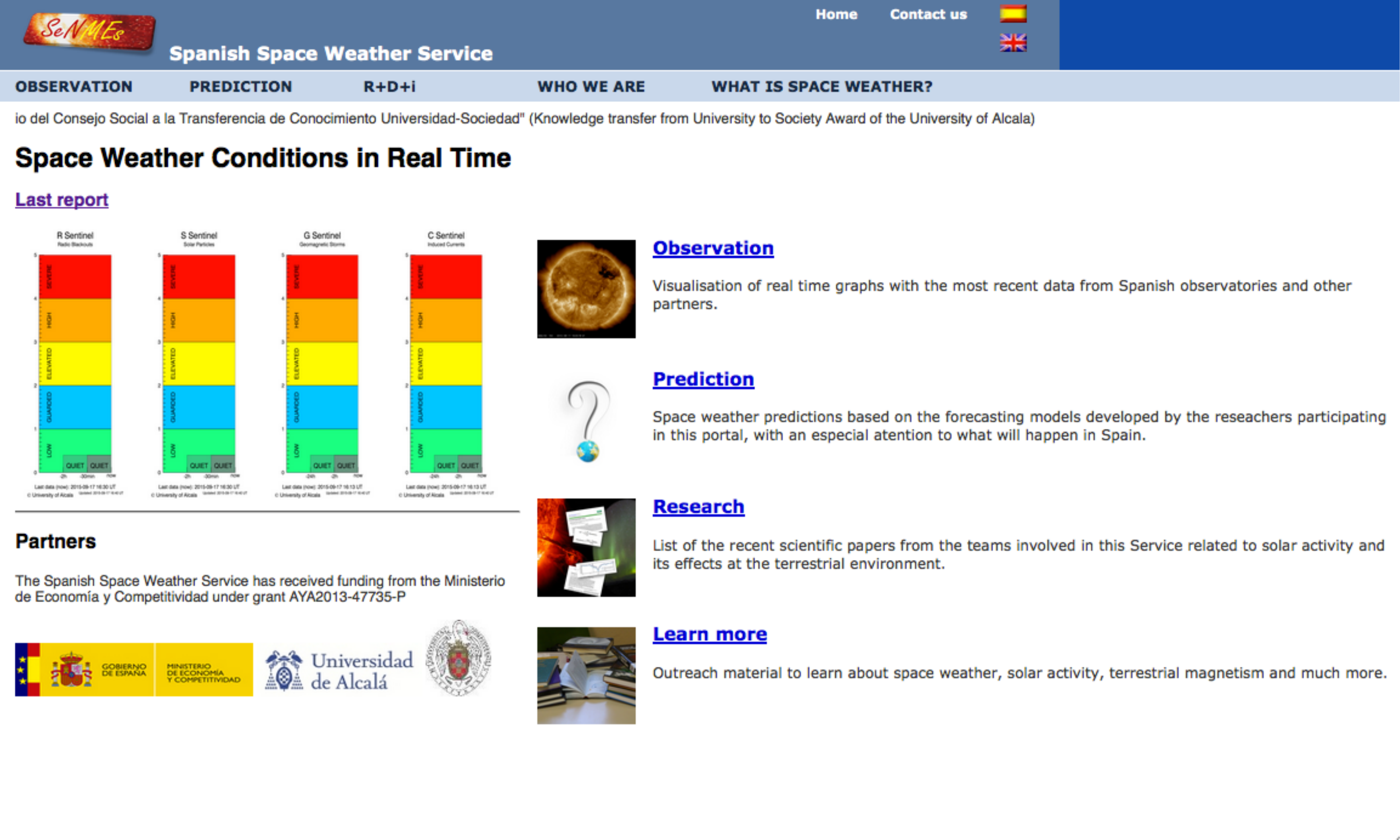}{cspm2015palacios1_fig1}{Home of the webpage SeNMEs. For colour version figures, please check the on-line version.}

\articlefigure{\figspath/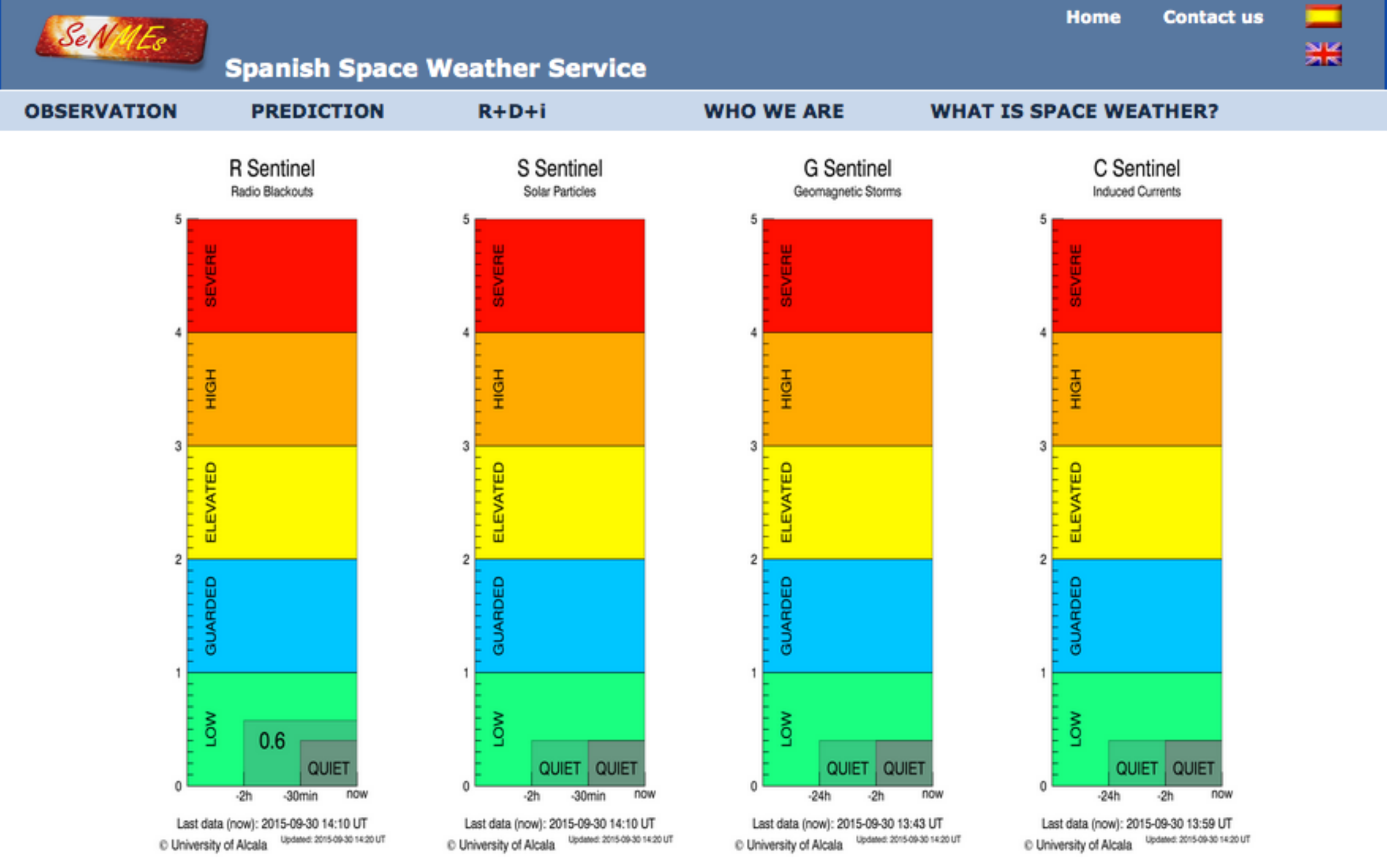}{cspm2015palacios1_fig2}{Risk scales of SeNMEs. For colour version figures, please check the on-line version.}

\subsection{Observations}

In the `Observations' webpage section we have summarized the Sun-Earth chain, along with the own research groups nowcasting products, displayed in Fig.~\ref{cspm2015palacios1_fig3}. These products are the Local Disturbance Index \emph{LDi{\~n}}, its derivative and the ionospheric map. The \emph{LDi{\~n}} is elaborated in real-time at UAH from SPT geomagnetic observatory data after removing quiet time values.
The ionospheric map shows Total Electron Content, TEC, over the Iberian Peninsula at date and time indicated on the image. TEC has been obtained by processing GPS/GLONASS observations from RINEX files from GNSS stations belonging to EUREF Permanent Network and International GPS Service, IGS, networks. TEC calibration technique provided by Professor Luigi Ciraolo has been used \citep{Ciraolo2007}. This technique assumes the thin shell model of the ionosphere to obtain TEC at the Ionospheric Pierce Points (IPPs). The calibrated TECs at IPPs have been interpolated using Ordinary Kriging method to obtain the maps. The TEC unit (TECu) is 10$^{16}$~e$^{-}$m$^{-2}$.

These maps are daily elaborated by Grupo de Estudios Ionosf{\'e}ricos y T{\'e}cnicas de Posicionamiento Global por Sat{\'e}lite (GNSS) from Universidad Complutense de Madrid.

\articlefigure{\figspath/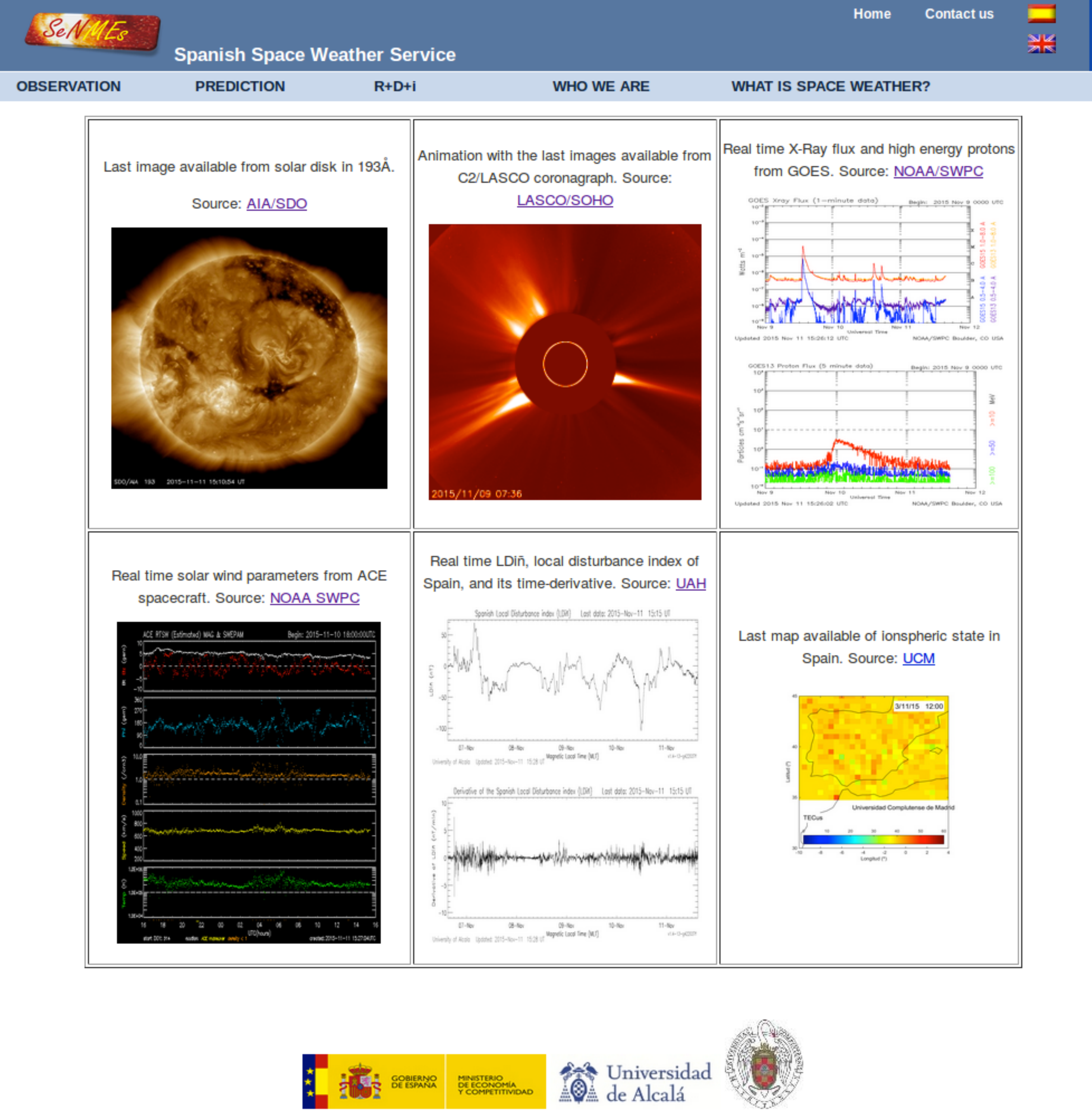}{cspm2015palacios1_fig3}{SeNMEs observation section, indicating the products developed by the SeNMEs team: \emph{LDi{\~n}}, its derivative, and ionospheric map are in the two last bottom panels. For colour version figures, please check the on-line version.}

\subsection{Predictions}

There are several prediction services for geomagnetic storms. The first one is about the onset time of geomagnetic storms, with a colourbar flag in \url{www.spaceweather.es}.The warning service consists on a true-false function:  `true' (red) when a $\Delta Dst$ larger than 50 nT in an hour is expected, `false' (green) otherwise. Data cadence is 1~minute.
The rationale of this prediction is in \citet{Saiz2008}. The potential users of this product include operators of power companies or companies involved in any other system affected by geomagnetically induced currents.

The second is the prediction of the recovery phase of the storm, i.e., the time for the magnetosphere to recover. The product provides an estimation of the time remaining for the magnetospheric recovery after an intense geomagnetic storm and the evolution of the \emph{Dst} index until quiet time. The basis of this empirical model is in \citet{Aguado2010}. Served data output cadence is 30~minutes. The potential users of this product include operators of any company involved in systems affected by disturbances in the terrestrial environment related to electric or magnetic fields.

\section{Case study: the geomagnetic storm on 22-23 June 2015}

This geomagnetic storm had basically two solar triggers: the CMEs from the anemone region AR 12371. These were halo-CMEs, the first one ejected on June 21, 02:36 UT, with a plane-of-the sky-speed of 1100~km~s$^{-1}$,  associated to a M2.7-class flare; and the second was ejected on June 22, 18:36 UT, with a plane-of-sky speed of 1300~km~s$^{-1}$, and associated to a M6.5-class flare. Mid-high energy proton flux is enhanced at that time, but not associated with flares. The previous CME on June 19, mentioned in the report, may also have played a role. Both CME speeds correspond with the time on the large feature in \emph{LDi{\~n}} on 22 and the drop on mid 23. The anemone region is an active region developed in a coronal hole \citep[see references in][]{Cid2014}. In this case it is very sheared and chrysanthemum-shaped, showing very fine structure. 

The report at that time was the following: \emph{Today June 23, there are 5 active regions (AR). Flare energy levels remain low, but some M-class flares on June 21.
There was a partial halo from the South on June 19 at 07:30 UT, with a plane-of-the-sky (POS) speed of 400~km~s$^{-1}$, due to an eruption from a large filament in that hemisphere. On June 21 there was a weak full halo CME at 02:36 UT, with a POS speed of about 1200~km~s$^{-1}$, from the anemone region AR 12371. There were also some weak ejecta from the SE (backsided); and SW limb, from AR 12367. Flares from AR 12371 from June 22, 01 UT on were registered. There is some probability of a full halo CME to be ejected at 06 UT on. Other full halo CME was ejected from AR 12367 at 18:36 UT with a maximum POS speed of 1300~km~s$^{-1}$, associated with a M6.5-class flare.
On June 23, at 00 UT a wide limb CME was ejected with a POS speed of 800~km~s$^{-1}$ from AR 12367. After that, a filament eruption from the northern solar hemisphere at 06:55 UT will probably lead to a wide and slow CME. LASCO data are not available for this time.
The particle event may be related to AR12367 flares. The proton peak observed at 18 UT may be due to a M6.5 flare.
There is a central coronal hole (CH) and large filaments on both hemispheres.
Solar wind velocity is above 550~km~s$^{-1}$ and interplanetary magnetic field strength is about 10~nT. These values corresponds to disturbed solar wind.
Both, \emph{LDi{\~n}} and its temporal derivative have presented important disturbances, as forecasted in the last report issued by SeNMEs. The derivative passed 50~nT~min$^{-1}$, which corresponds to a severe event according to C-scale. According to scale G the disturbance was less important because \emph{LDi{\~n}} did not pass the threshold of --150~nT.
The arrival of the halo CME on June 22 is expected from the first hours of June 24. The CME on June 23 may also reach the terrestrial environment. Because the preconditioning of the magnetosphere, important disturbances are expected in the next 24 hours.} The G-~and C-scale maximum values and the ionospheric state are shown in Fig.~\ref{cspm2015palacios1_fig4} and   local geomagnetic values, in Fig.~\ref{cspm2015palacios1_fig5}.

\articlefiguretwo{\figspath/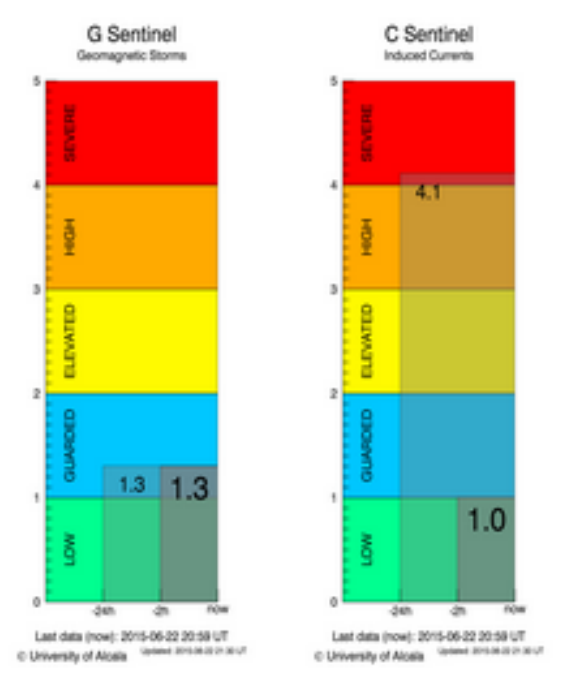}{\figspath/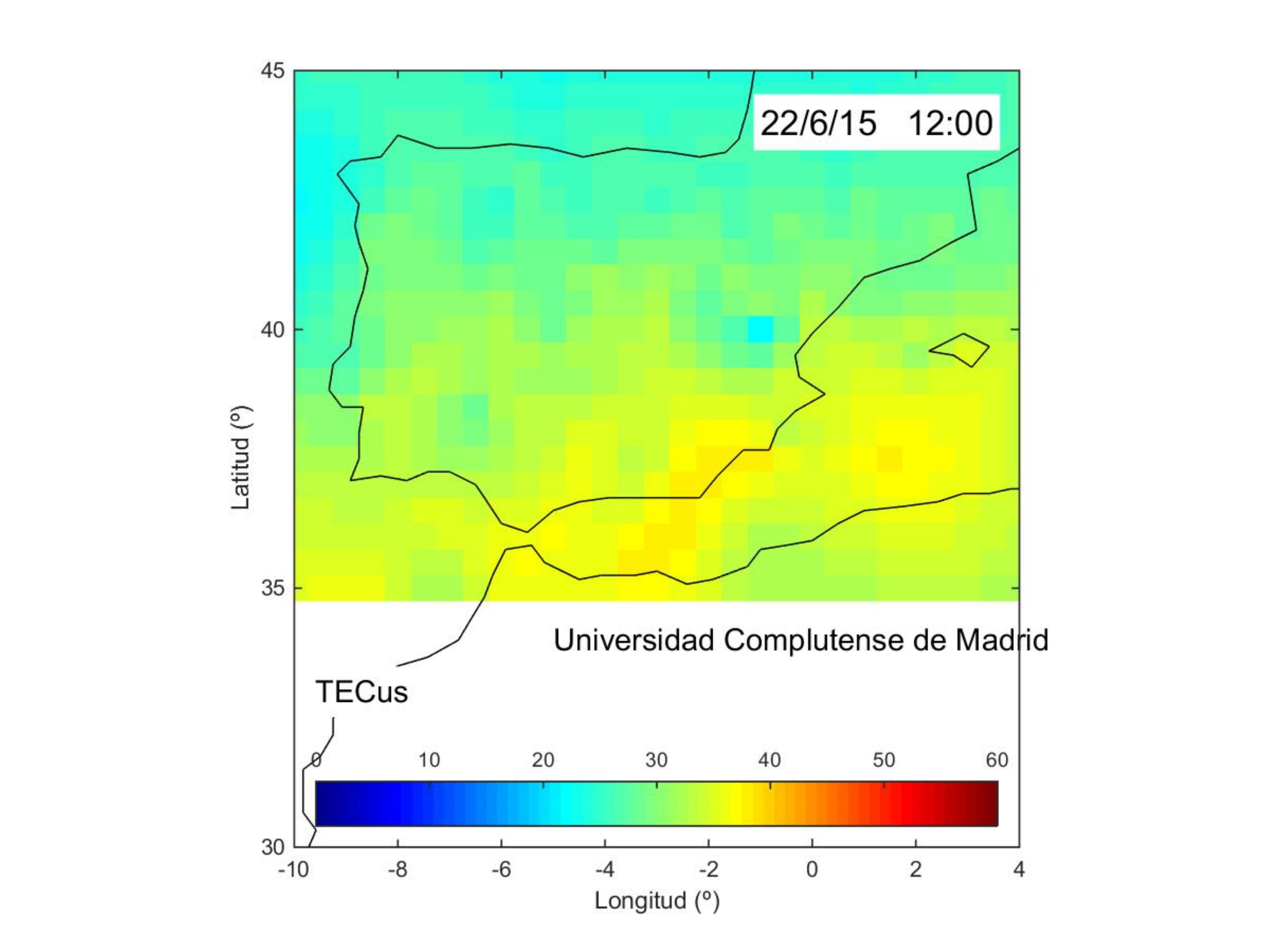}{cspm2015palacios1_fig4}{Two summarizing figures of the effects of the storm on Earth. \emph{Left:} G-~and C-scales for this storm. \emph{Right:}  Ionospheric map over the Iberian Peninsula. For colour version figures, please check the on-line version.}

\articlefigure{\figspath/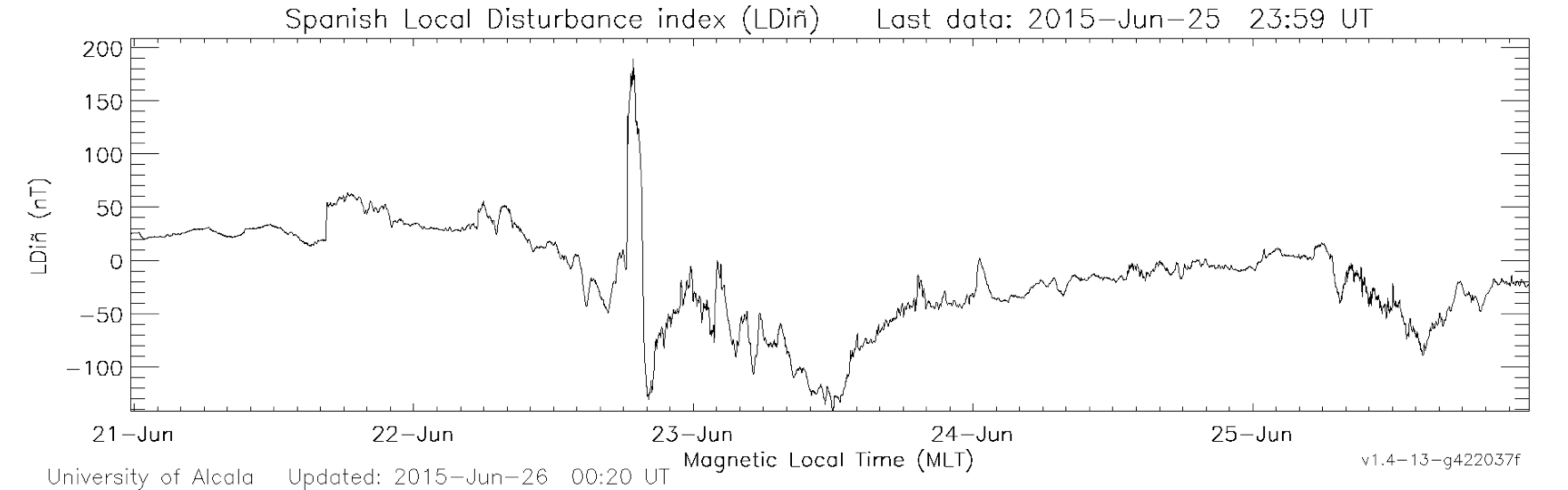}{cspm2015palacios1_fig5}{\emph{LDi{\~n}} plot of the geomagnetic storm.}


\acknowledgements The authors want to acknowledge and thank IGN, NOAA, IGS, EUREF, ESAC, SDO/AIA and LASCO/SOHO Data Science Centers and Teams, also to ACE and GOES teams. Special thanks to the UAH travel grant and CSPM for financial support, and MINECO project AYA2013-47735P. Thanks also to Helioviewer for data visualization. Thanks to Prof. Ciraolo for the TEC calibration technique.

\bibliography{cspm2015_palacios1}  

\end{document}